\documentclass[]{mn2e}
\usepackage{graphicx}
\begin{document}

\title{Cosmic Ray Scattering in Compressible Turbulence}
\author[A. Lazarian \& A. Beresnyak]{A.~Lazarian,$^1$
A.~Beresnyak,$^1$ \\
$^1$University of Wisconsin-Madison, Dept. of Astronomy}

\maketitle

\begin{abstract}
We study the scattering of low-energy Cosmic Rays (CRs) in a turbulent,
compressive MHD fluid. We show that compressible MHD modes -- fast or
slow waves with wave lengths smaller than CR mean free paths induce
cyclotron instability in CRs.  The instability feeds the new
small-scale Alfv\'enic wave component with wave vectors mostly along
magnetic field, which is not a part of the MHD turbulence cascade.
This new component gives feedback on the instability through decreasing
the CR mean free path. We show that the ambient turbulence fully suppresses
the instability at large scales, while wave steepening constrains the
amplitude of the waves at small scales. We provide the energy spectrum of
the plane-parallel Alfv\'enic component and calculate mean free paths
of CRs as a function of their energy.  We find that for the typical
parameters of turbulence in the interstellar medium and in the
intercluster medium the new Alfv\'enic component provides the scattering
of the low energy CRs that exceeds the direct resonance scattering by
MHD modes. This solves the problem of insufficient scattering of
low-energy CRs in the turbulent interstellar or intracluster medium that
was reported in the literature.
\end{abstract}


\section{Introduction}

Cosmic rays (CRs) and magnetic fields are essential components for many
astrophysical ecosystems, including galaxies and clusters of galaxies
(see Schlickeiser 2002). In many instances, e.g. Milky Way, the
pressure of CRs and magnetic fields is larger than the gas
pressure. As a rule, astrophysical magnetic fields are frozen in turbulent
plasma and move together with it. As a result, CRs interacting
with turbulent magnetic fields get scattered and accelerated (see
Melrose 1968, Schlickeiser 2002). 
  
The magnetohydrodynamic (MHD) approximation is widely used to describe
the actual magnetized plasma turbulence over scales that are much
larger than both the mean free path of the particles and their Larmor
radius (see Kulsrud 2004). The theory of MHD turbulence has become
testable recently due to numerical simulations (see Biskamp 2003) and
this provided reliable foundations for describing turbulence-CRs
interactions. The simulations (see Cho \& Lazarian 2005 and
ref. therein) confirmed the prediction of magnetized Alfv\'enic eddies
being elongated along magnetic field (see Shebalin, Matthaeus \&
Montgomery 1983, Higdon 1984) and provided results consistent with the
quantitative relations for the degree of eddy
elongation obtained  in Goldreich \& Sridhar (1995, henceforth GS95).

Scattering of CRs is an essential part of both CR propagation modes
and models of CR acceleration. Efficient scattering is usually
postulated (see Schlickeiser 2002), which ensures high degrees of
coupling of CRs and magnetized plasma. In addition, efficient
scattering provides appreciable second order Fermi acceleration and
enables the return of CRs into the shock to ensure the first order Fermi
acceleration.

This corner stone of CR physics has been challenged recently when
it became clear that Alfv\'enic eddies are stretched along magnetic
field direction.  As the interaction between CRs and such elongated
eddies is weak (see
discussion in Lerche \& Schlickeiser 2001), this resulted in the
prediction of long mean free paths for Milky Way CRs (Chandran 2000,
Yan \& Lazarian 2002, henceforth YL02). YL02 and Yan \& Lazarian
(2004) attempted to remedy the situation by appealing to CR
scattering by isotropic sound-like fast modes. However,
plasma-dependent damping of fast modes made the scattering very
different in different parts of the interstellar medium. Is such a
radical change of the CR scattering picture absolutely necessary?

We note, that the problem of CR scattering goes well beyond the
Milky Way physics.  Brunetti (2006) discussed the implications of
suppressed CR scattering on the acceleration of CRs in the clusters of
galaxies.  Is there any process through which scattering by fast modes can
provide high efficiency of CR scattering?  Below we consider such a
process that is related to CR feedback on MHD turbulence. We show
that compression of CRs induces instability that results in the
generation of modes that are parallel to the magnetic field. Such
modes that are also frequently referred to as slab modes have been long
employed in the models of CR propagation (see, e.g., Jokipii 1966),
their origin, however, was somewhat mysterious.
This paper provides a physically motivated mechanism for the generation
of slab modes and quantify the efficiency of their generation.

In what follows we discuss the properties of
compressible MHD turbulence in \S~2. We describe the kinetic
instability that develops in CRs when the magnetic field is compressed
on scales less than the CR mean free path in \S~3.
We consider the non-linear saturation of the instability in
\S~4 and its large-scale cut-off that follows from the
interaction of the instability waves with the ambient turbulence in
\S~5. The feedback of CRs to compressible turbulence is considered in
\S~6. The implications of our work for various ISM phases and intra-cluster
medium (ICM) are considered in \S~7.
The short summary of this work is presented in \S~8.

\section{Compressible MHD Turbulence}

In this section we provide a summary of the current knowledge of
compressible MHD turbulence that we appeal to in our work. As we
mentioned earlier, we describe an alternative mechanism for the scattering
of low-energy CRs that provides rather small mean free paths for
CRs. At scales larger than the mean free path CRs are well coupled to the
flow of the thermal plasma.  Therefore to identify new effects we
concentrate on the MHD fluctuations on scales equal or smaller than
the mean free path\footnote{A notable exception from this rule, MHD shocks,
  relevant for CR acceleration, is not considered in this paper.}.

The GS95 model quantifies turbulence anisotropy, introducing the
so-called {\it critical balance} relation $k_{\|}\sim k_{\bot}^{2/3}$,
where $k_{\|}$ and $k_{\bot}$ correspond to, respectively, the parallel
and perpendicular wavenumber of eddies measured in respect to the {\it
  local} magnetic field. This scaling is true for both Alfv\'enic and
pseudo-Alfv\'enic motions, which are the incompressible limit of the
slow modes. GS95 and later Lithwick \& Goldreich (2001) argued that
the slow modes are passively advected by the Alfv\'en modes, while the
energy exchange between the modes is small. Numerical studies (Cho \&
Vishniac 2000, Maron \& Goldreich 2001, Cho, Lazarian \& Vishniac
2002, M\"uller, Biskamp \& Grappin 2003) provided reasonable support for the critical
balance condition\footnote{Note that the GS95 prediction of the Kolmogorov
  spectrum $-5/3$ spectral index stayed more controversial, as Maron
  \& Goldreich (2001) reported the spectrum closer to $-3/2$. While
  the ongoing work (see Muller \& Grappin 2005, Boldyrev 2005, 2006,
  Beresnyak \& Lazarian 2006) attempts to improve our understanding of
  Alfv\'enic turbulence, we shall use the original GS95 scalings for
  the sake of simplicity of our presentation.}.
We introduce the outer scale $L_A$ and use GS95 scaling in the form
\begin{equation}
k_{\|}\sim k_{\bot}^{2/3}L_A^{-1/3},~~~ \delta v\sim v_A (k_{\bot}L_A)^{1/3},
\label{GS95}
\end{equation}

We feel that GS95 provides a good starting point for studies of mildly
compressible, and even supersonic MHD turbulence.  Indeed, numerical
calculations in Cho \& Lazarian (2002, 2003, henceforth CL02, CL03)
showed that scalings of the slow and Alfv\'en modes in compressible
MHD are similar to their scalings in the incompressible case. The fast
mode perturbations, on the other hand, are found to be mostly
isotropic with a power-law index of about $-3/2$ (see CL02), which
is the index of so-called acoustic wave turbulence. The coupling of
the fast and Alfv\'en modes was shown to be weak, which allows separate
studies of the corresponding cascades provided that the Alfv\'enic
turbulence is strong, i.e. it evolved to develop the critical balance.

In the following treatment we will be primarily interested in two
manifestations of compressible MHD turbulence, one of which is
perturbations of the {\it magnitude} of the magnetic field and the
other is wave damping through cascading by the ambient Alfv\'enic
turbulence. Let us briefly discuss how much of our results depend on
the adopted model of MHD turbulence.

As we are dealing with very small perturbations down in the inertial
range, Alfv\'enic mode, having magnetic field perturbations
perpendicular to the local magnetic field, has very little effect on
the magnetic field magnitude, therefore, as far as the magnetic field
compression is concerned, we are dealing with compressible
modes. We introduce the power-law energy spectrum for the velocity
perturbations of the compressible modes, $E(k)\sim k^{-\beta}$, where
$\beta=5/3$ for the Kolmogorov-type, and $3/2$ for the acoustic type
spectrum. Such a spectrum translates into velocities at the scale
of $l$, which is $\delta v_{l}\sim v_A (l/L)^\mu$, where
$\mu=(\beta-1)/2$ and $L$ is not necessarily the injection scale for MHD
turbulence, but a quantity which contains information on both the injection
scale and the efficacy of driving.  This scaling might not be valid up
to scales as large as $L$.  If both sonic and Alfv\'enic Mach numbers
are around unity at the injection, we expect $L$ to be of the order of
the actual injection scale. Efficacy of driving depend on its type,
such as mostly solenoidal, or mostly compressive (supernova shocks,
etc.). A precise estimate of the parameter $L$ requires either detailed
knowledge of the physics of driving, or direct measurement of
the compressive fluctuation intensity somewhere in the inertial range.

The Alfv\'en wave damping by Alfv\'enic turbulence we use in \S~4
assumes GS95 scaling and anisotropy with some outer scale $L_A$. This
scale might correspond to the isotropic injection of energy at scale
$L$ and the injection velocity of $v_A$, i.e. the Alfv\'en Mach number
$M_A\equiv (\delta v/v_A)=1$.  This model can be easily generalized
for both $M_A>1$ and $M_A<1$ at the injection.  Indeed, if $M_A>1$,
instead of the driving scale $L_D$ for $L_A$ one can use the scale at
which the turbulent velocity gets equal to $v_A$.  For $M_A\gg 1$
magnetic fields are not dynamically important at large scales and the
turbulence follows the Kolmogorov cascade $v_l\sim l^{1/3}$ over the
range of scales $[L_D, L_A]$. This provides $L_A\sim L_D M_A^{-3}$.
If $M_A<1$, the turbulence obeys GS95 scaling (also called ``strong''
MHD turbulence) not from the scale $L_D$, but from a smaller scale $l'
\sim L_D M_A^{2}$ (Lazarian \& Vishniac 1999), while in the range
$[L_D, l']$ the turbulence is ``weak''.  The velocity at scale $l'$ is
expressed as $v_{l'}\sim v_A M_A^{2}$, so that the ``effective'' value of
$L_A$ will be $L_A=L_DM_A^{-4}$.

All in all, given the strength and the nature of driving in a
particular astrophysical environment one may estimate the two
parameters, $L$ and $L_A$ that determine the velocity perturbations of
the compressible and Alfv\'enic mode at small scales.

In \S~3 we deal with the fluctuations of the magnetic field magnitude
squared.
The normalized amplitude of these fluctuations denoted as $A$ will
depend on the plasma $\beta$ which is the ratio of the gas pressure to
the magnetic field pressure. In high-$\beta$ plasmas $B^2$ will be
perturbed mostly by slow waves (CL03) and the value of $A$
will be equal to $2(\delta v/v_A)\sin\theta$, where $\theta$ is the
angle between the wave vector and the magnetic field. In the inertial
range of strong turbulence the slow mode exhibits the same anisotropy as
the Alfv\'enic mode.  Therefore $\theta$ is close to $90^o$, so we can
disregard the angular factor, i.e.  $A=2\delta v/v_A$. The perturbation
made by the fast mode in high beta plasmas is smaller by a factor of
$v_A/c_s$, where $c_s$ is sound velocity.  In low-$\beta$ plasmas this
situation is reversed, with the slow mode only marginally perturbing $B$,
but we may use the same expression $A=2(\delta v/v_A)\sin\theta$
for the fast mode. We estimate the angular factor as of the order of
unity, since fast modes are almost isotropic (CL02). In other words,
the expression $A=2\delta v/v_A$ is interpreted as the compression factor
for the {\it most compressive mode}\, which is the slow wave in
high-$\beta$ plasmas and the fast wave in low-$\beta$ plasmas.

The lower limits to the scales we described are determined by the
damping of MHD modes. While the Alfv\'enic mode in fully ionized media
is supposed to be damped at the thermal Larmor radius, the compressive
modes are damped by more efficient collisional and collisionless damping
(Ginzburg 1961, Barnes \& Scargle 1973, YL04). In this paper we introduce
the value of the most compressive mode cut-off scale as $l_{\rm cut}$.
 
\section{Instability of Compressed CRs}

It is obvious that at scales less than their mean free path CRs can be
treated as a collisionless fluid. Particles in the collisionless fluid
preserve the adiabatic invariant $p^2_\perp/B$, where $B$ is the the
magnetic field strength and $p_\perp$ is the momentum perpendicular to
the magnetic field. MHD compressive modes change the magnitude of $B$
so that the
distribution in momentum space becomes anisotropic (see Chew,
Goldberger \& Low, 1956).

Such a distribution is subject to a number of instabilities, some of
which are hydrodynamic, i.e. involve the change of the entire
distribution function, while others are kinetic, i.e. involve a chance
in a fraction of particles, that is resonant with a particular
wave-mode. Well-known examples of hydrodynamic instabilities are
firehose and mirror instabilities (see e.g. Mikhailovskii,
1975). Hydrodynamic instabilities are typically fast with the largest
wavenumber growing almost as fast as the gyrofrequency, but have a
threshold, i.e. small deviations from isotropy do not induce
instability.

While compressive motions can generally induce rather large changes in
the magnitude of $B$ on scales at the injection scale of turbulence
$L$, in \S3 we estimate mean free path and show that it is much
smaller than $L$.  Therefore the compressions of the magnetic field we
deal with are too small to induce hydrodynamic instabilities.

It has been well known that the momentum distribution functions with
$p_x=p_y>p_z$, are subject to kinetic instability called gyroresonance
instability (Sagdeev \& Shafranov, 1961, Mikhailovskii, 1975, Gary, 1993).
This instability received less attention than its hydrodynamic counterparts.
However, it is pretty fast for a power-law distribution of
CRs, as we demonstrate below.

For a power-law distribution of CRs the growth rate of the
cosmic-ray-Alfv\'en gyroresonance instability (henceforth GI) can be
estimated as (see Appendix):
\begin{equation}
\gamma_{\rm CR}(k_{\|}) = \pm \omega_{pi} \frac{n_{\rm CR}(p>m\omega_B/k_{\|})}{n}AQ,
\label{stream}
\end{equation}
where $n_{\rm CR}(p>m\Omega/k_\|)$ is the number density of CRs with
momentum larger than the minimal resonant momentum for a wave vector
value of $k_\|$, $m$ is the proton mass, $n$ is the density of the thermal
plasma, $\omega_{pi}$ is the ion plasma frequency. $Q$ is a numerical
factor, defined in the Appendix.  The $\pm$ sign corresponds to the
two MHD modes. We shall concentrate on the Alfv\'en mode, corresponding to
the plus sign, as those are less subjected to linear damping (see
\S~2, \S~6). As we will demonstrate in the next chapter, when anisotropy is
created by compressive turbulence, the anisotropy factor
$A=(p_\perp-p_\|)/p_\|$
will be small and will change its sign on the scale of the mean free path,
depending on two competitive mechanisms -- scattering which tends to
isotropize momentum distribution, and magnetic field compression
which tends to make it oblate or prolate. 

We assumed that the unperturbed distribution of CRs is isotropic and follows
a power law i.e.  $F_0 \sim p^{-\alpha-2}$ where $\alpha$ is
conveniently defined as the power-law index for a one-dimensional
distribution (or particle density). For example, around the Earth $\alpha
\sim 2.6$ up to the energies of $10^{14}$~eV. Note, that in order for the
total energy to converge at high energies, $\alpha$ should be larger
than $2$.

The expression for the instability rate, assuming $A=2\delta v/v_A$ (see \S~2),
could be written as
\begin{equation}
\gamma_{\rm CR}(r_p)=\frac{\delta v}{L_i}\left(\frac{r_p}{r_0}\right)^{-\alpha +1},
\label{instab_main}
\end{equation}
where $r_p$ is a Larmor radius of a CR resonant with a particular wave
vector $k_\|=m\Omega/p$, $r_0$ is the 1~GeV proton Larmor radius and
\begin{equation} 
L_i=3.7\cdot10^{-7}\frac{1}{Q} 
\left(\frac{B}{5\cdot10^{-6}~\mbox{G}}\right)
\left(\frac{4\cdot10^{-10}~\mbox{cm}^{-3}}{n_{\rm CR}(r_p>r_0)}\right)~{\rm pc}.
\label{lili}
\end{equation}

\section{Non-linear Suppression and Saturation}

We introduce the CR mean free path $\lambda$ below which CRs could be
treated as collisionless and the instability described in \S~3 is
active. In the absence of other scattering processes the CRs
are scattered by the slab-type motions generated by the instability
above. Let us estimate $\lambda$ following Longair (1994).  If the
change of magnetic field direction is $\phi\sim \delta B/B$ the
scattering that is a random walk requires $N\sim 1/\phi^2$ interaction
and
\begin{equation}
\lambda\sim Nr_p\sim r_p/\phi^2\sim r_p B^2/(\delta B)^2,
\label{mean_path}
\end{equation}
where we designated $\delta B$ as the magnetic field perturbation
pertaining to a particular wavenumber, i.e.  $\delta B^2\approx
E(k)k$. We can consider $\delta B$ as a function of either $k$ or the
resonant Larmor radius $r_p$ (see Longair 1994). As the instability
grows, $\delta B$ will grow, which reduces $\lambda$. On the other
hand, it is the mean free path $\lambda$ which determines the scale at
which compressions of the magnetic field are important. This can be
understood as follows: the CR distribution ``remembers'' the perturbed
value of the magnetic field and its anisotropy only during the
time the typical particle travels its mean free path. Once particles
scatter significantly, the anisotropy of the distribution is
effectively ``reset''. As a result only low amplitude motions on
scales less than $\lambda$ excite the instability, or in other words,
the degree of anisotropy $A$ is determined by the {\it local}
perturbation of the magnetic field on the the scale $\lambda$. We call this
process {\it non-linear suppression}.

The instability grows as $d(\delta B^2)/dt=\gamma_{\rm CR} (\delta B^2)$
where the injection of energy is happening at the scale of the mean
free path, i.e.
\begin{equation}
\gamma_{\rm CR}\approx\frac{v_{A}}{L_i}
\left(\frac{r_p}{L}\right)^{\mu}\left(\frac{\delta B}{B}\right)^{-2\mu}
\left(\frac{r_p}{r_0}\right)^{-\alpha+1},
\label{instab_limit}
\end{equation}
where eqs. (\ref{instab_main}) and (\ref{mean_path}) were used.  We see
that according to the above equations $\delta B$ perturbations will grow
as $t^{1/2\mu}$ thus reducing $\lambda$ virtually to $r_p$. In other
words, the non-linear suppression is not able to constrain the
development of instability and we have to consider other non-linear
processes, such as wave steepening.

Steepening does not occur for a monochromatic circularly polarized
Alfv\'en wave as the amplitude of the magnetic field stays the same.
However, for
a collection of waves with different wavelengths the amplitude of
the magnetic field fluctuates in space and time and therefore the
steepening effect is present. The steepening rate can be estimated as
\begin{equation}
 \gamma_{\rm steep} \approx -(\delta B/B)^2 k_{\|} v_A,
\label{steep}
\end{equation}
where the ``-'' sign reflects the fact, that steepening damps the
instability.

By comparing (\ref{instab_limit}) and (\ref{steep}) we get the
equilibrium or saturated amplitude of the instability-induced
perturbations
\begin{equation}
\frac{\delta B}{B}\approx
\frac{r_0^{1/2}}{L_i^{1/(2\mu+2)} L_{\phantom{i}}^{\mu/(2\mu+2)}}
\left(\frac{r_p}{r_0}\right)^{(\mu-\alpha+2)/(2\mu+2)},
\label{amplitude}
\end{equation}
which for $\alpha=2.6$ and $\mu=1/3$ produces a rather shallow spectrum
of perturbations, $E(k)\approx (\delta B)^2/k\sim k^{-0.8}$.

Combining Eqs. (\ref{mean_path}) and (\ref{amplitude}) one gets that
the energy of the slab modes at $k_\|\sim r_p$ is supplied from
the compressions at scale
\begin{equation}
\lambda \approx L_i^{1/(\mu+1)} L_{\phantom{i}}^{\mu/(\mu+1)}
\left(\frac{r_p}{r_0}\right)^{(\alpha-1)/(\mu+1)}.
\label{lambda_suppl}
\end{equation}

So far we assumed that the turbulent compressible motions are not
damped. This is a good approximation until $\lambda$ is larger than
the compressive mode cutoff scale $l_{\rm cut}$. If, on the other hand,
$l_{\rm cut}>\lambda$ the compression for the instability
is supplied from the eddies at the damping scale, namely,
$\delta v/v_A\sim (l_{\rm cut}/L)^{1/3}(\lambda/l_{\rm cut})$.
The modification of our formulae is self-evident. Instead
of eq. (\ref{amplitude}) one gets
\begin{equation}
\frac{\delta B}{B}\approx\left(\frac{r_0^{1/2}}{L_i^{1/4}L_{\phantom{i}}^{\mu/4}l_{\rm cut}^{(1-\mu)/4}}\right)^{1/4}
\left(\frac{r_p}{r_0}\right)^{(3-\alpha)/4},
\label{amplitude2}
\end{equation}
which, for the same value of $\alpha=2.6$, corresponds to a steeper
spectrum of $E(k)\sim k^{-1.2}$.

\begin{figure}
\includegraphics[height=100mm]{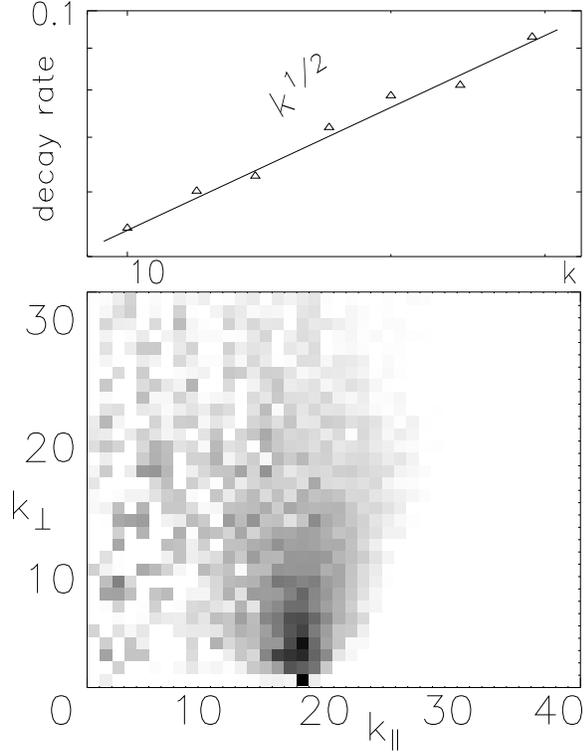}
\caption{Decorrelation of a plane, $k_\perp=0$ Alfv\'en wave by
  turbulence. Lower picture shows the energy density of a wave in
  cylindrical k-space. In this picture Alfv\'en waves were injected
  at $k_{\|}=17$.
  Wave energy is being transferred in the direction of $k_\perp$ axis,
  which is typical for decorrelation by MHD turbulence. 
  Wave decay was exponential after driving was switched off.
  Upper plot shows decay rate of the wave vs its wavenumber.
  A detailed discussion is presented in Beresnyak \& Lazarian (2006).  }
\end{figure}

\section{Damping by Alfv\'enic Turbulence}

The instability we considered in \S~3 has the largest growth rates for
the wave vector parallel to the field.  This is due to the fact, that
the phase of a resonance can be kept constant for a long time only if
$k_{\bot}$ is small. The ambient turbulence non-linearly damps the
instability through a process that is analogous to the suppression of
the streaming instability (YL02, Farmer \& Goldreich 2004).  In what
follows we find the lower limit on $k_{\bot}$ using the approach
similar to that in Farmer \& Goldreich (2004). We also provide the
results of numerical calculations that validate this approach.

For Alfv\'enic turbulence we adopt the GS95 scaling, (\ref{GS95}), which
reflects the tendency of eddies to get elongated along the magnetic
field. For the sake of simplicity we take the scale $L_A$, introduced
in \S 2, equal to the scale $L$. This is not necessarily true for any
astrophysical environment, however our formulae are trivially
generalized for the case of $L_A \neq L$.

Consider a wavepacket of Alfv\'en waves that moves nearly parallel to
the magnetic field with the dispersion of angles $\delta
k_{\bot}/k_{\|}\sim \theta_k$. The individual waves follow the local
direction of the magnetic field lines. As a result, the dispersion in
angles of the wave packet cannot be less than the dispersion of angles
due to the ambient Alfv\'enic turbulence, $\theta_k>\theta_{bk}$. The
latter for the GS95 model (see eq.~(\ref{GS95})) is $\theta_{bk}\sim
\delta B_k/B_0 \sim (k_{\bot}L)^{-1/3}$.  The modes with minimal
$\theta_k$ are the fastest growing ones.  As we establish below (see
Eq.~(\ref{turb})), they are the least damped.  Therefore for our
simplified treatment we shall limit our attention to the wavepackets
with resonant $k_{\|}^{-1}\sim r_p$ and $\theta_{bk}\sim
\theta_k$. One can determine the characteristic
perpendicular wavenumber $k_{\bot}\sim \delta k_{\bot}\sim
r_p^{-1}(r_p/ L)^{1/4}$ of the ``most parallel modes'' that are
created by streaming CRs.

The strong Alfv\'enic turbulence decorrelates the wave\-packet with
$k_{\bot}$ on the time scale of $v_{\bot} k_{\bot}$.  Thus using the
above expression for $k_{\bot}$ and Eq. (\ref{GS95}) we get
\begin{equation}
\gamma_{\rm turb}\sim -k_{\bot} v_{\bot} \sim -v_A k_\perp^{2/3} L^{-1/3} 
\sim - v_A r_p^{-1/2} L^{-1/2},
\label{turb}
\end{equation}
which, up to the ``-'' that we used to denote the damping nature of
the process, coincides with the damping rate obtained in Farmer \&
Goldreich (2004) and with the results of our numerical simulations
shown in Fig. 1.

A comparison between eqs. (\ref{instab_main}) and (\ref{turb}) indicates that for
the spectral index of CRs $\alpha >3/2$ the ambient Alfv\'enic turbulence
provides an upper limit on the scale of perturbations that arise from
compressible-induced instabilities even without accounting for
nonlinear suppression.

If we use nonlinear suppression, the critical scale can be obtained by
using eqs.~(\ref{instab_limit}), (\ref{amplitude}) and (\ref{turb}).
For $\alpha>5/3$ our instability is damped for all scales, larger than
\begin{equation}
r_{p,{\rm crit}}\approx r_0\left(L_{\phantom{i}}^{1-\mu}r_0^{\mu+1}L_i^{-2}\right)^{1/(2\alpha-\mu-3)}.
\label{crit_rad}
\end{equation}
Therefore the spectrum of plane Alfv\'en waves given by
eq.~(\ref{amplitude}) will protrude from $r_{p,{\rm crit}}$ down to
$r_{p,{\rm min}}$ which corresponds to minimum energies of CRs.

\section{Feedback on Compressible Turbulence}

Linear instability theory does not describe the energy transfer
between particles and CRs. Moreover, in calculating linear response
we have not included the CR input into the real part of the
dielectric tensor. Speaking of our description of CR scattering
by waves, we can consider each resonant scattering event in the rest frame
of the wave where it is a pure pitch-angle scattering, i.e.
there is no energy transferred between the particle and the wave.
Basically, the model of linear instability and its nonlinear
suppression and saturation developed in \S 4 is unable
to fully describe the transfer of energy from compressible modes to
CRs and from CRs to high-frequency waves which are finally dissipated
by steepening. 

We would like to note that the transfer of energy is, in principle,
might be very important for the particular models of turbulence we are
using in \S 2. They are so-called {\it Kolmogorov-type} models
\footnote{This very general concept should not be confused with
a particular Kolmogorov theory for incompressible strong
Navier-Stokes turbulence} in which the turbulence is initiated
at large scales, is being local in k-space and is characterized
by the flow of energy from large to small scales\footnote{In some
cases this is a flow from small to large scales, as in Lengmuir
wave turbulence (see, e.g., Kaplan, Tsytovich, 1973).}.
Both slow wave passive transfer by strong Alfv\'enic
turbulence and weak fast wave turbulence models we use in \S~2 are
of this type. When energy is being drained out of such turbulence
it does not any longer follow universal power-law scalings we described
in \S 2. In this section we will estimate the transfer
of energy from compressible motions and how it modifies the spectra
of turbulence and processes described in \S 4.

Let us first figure out the processes of transfer of energy from
CRs to waves. As we noted before, there is no transfer of energy
in the frame moving with the wave. Therefore, if we have waves
moving in only one direction, particles will pitch-scatter and
establish the drift equilibrium with such waves. The drift velocity
will be equal to the Alfv\'en velocity. This is the case of the
well-known streaming instability if we can neglect the wave damping.
In our case, however, we have waves moving in both directions, so
the particle can lose or gain energy in the lab frame by multiple
scattering. If each scattering event changes pitch angle by
$\phi\sim \delta B/B$ (c.f. eq (5)) the energy is changed
by $2p_\perp V_A \delta B/B$. This occurs at a Larmor frequency,
i.e. $c/r_p$. Now, starting with two above equations we can talk
about two processes, one of which is the well-known
diffusion in energy space, described by the diffusion coefficient
$D_{pp}$ estimated as

\begin{equation}
D_{pp}\approx \frac{p^2v_A^2}{r_pc} \left(\frac{\delta B}{B}\right)^2.
\end{equation}

This process is mostly describe the particle-wave equilibrium state
where energy is slowly redistributed between particles. Suppose, however,
that we efficiently drain energy from the waves so that the situation
become non-equilibrium. In this case we talk about the second process
which is a directed transfer of energy from particles to waves. The rate
of this process depends on the degree of non-equilibrium, but generally
cannot exceed the inverse of the minimum time at which particles can lose
the energy available from anisotropy, i.e.

\begin{equation}
\gamma_{\rm ex}=1/\tau_{\rm ex,min}\approx \frac{v_A}{r_p} \frac{\delta B}{B}\frac{1}{A}.
\end{equation}


This rate is actually $B/(\delta B A)$ times higher than a steepening
rate. More careful estimate show that the volumetric rate of the
energy exchange between particles and waves divided by the volumetric
steepening energy rate is equal to $B/(2\delta B AQ)$ which is much
larger then unity. Therefore we can conclude that if the energy
is drained from waves at a steepening rate, it will be efficiently
resupplied by CRs.

The anisotropy of CR distribution is, in turn, supplied by the
compressive perturbations. CR pressure is usually neglected in the
dispersion relation for the compressive perturbations. However we know
that it is there as CRs and the perturbation are connected by the magnetic field.
When a fraction of the CR pressure is lost due to the process described above,
some of the perturbation energy is lost too. This provides a new mechanism
for the damping of compressive perturbations in a medium with CRs.

As we noted previously, each particular scale of compressive motions
$\lambda$ transfers energy to a smaller scale of motions at
$r_p$. Also, we know that the flow of energy in Kolmogorov-type
turbulence is a constant. The steepening rate for the reasonable
range of parameters is, however, growing rather fast, as we show in the next
section. This could lead to a model where compressive motions are
fully damped at a particular scale $\lambda_{\rm fb}$ where Kolmogorov energy
transfer rate
is equal to the steepening rate at the scale $r_p(\lambda_{\rm fb})$.
More careful consideration, however,
shows a different picture. At the end of \S 4 we showed that scales larger
than $\lambda$ could also provide some compression for the CRs. If this
compression is higher than the one produced at $\lambda$, it will also
provide higher energy rate. Therefore, the compressive motions at scales,
smaller than our ``feedback damping'' scale are not damped fully, but
rather to the state where they provide marginal compression. It
is easy to see, that this correspond to the $\delta v \sim \lambda$
law, or spectrum of turbulence $E^{-3}$. This spectrum of compressive
motions will protrude from the scale $\lambda_{\rm fb}$ down to scale
$l_{\rm cut}$. The amplitudes of the high-frequency slab Alfv\'enic motions,
corresponding to these and lower scales will be determined by formula (10). 
The following handy expression and equations from \S 4 could be used to
derive $\lambda_{\rm fb}$:

\begin{equation}
\frac{\lambda_{\rm fb}}{L}\approx\left(\frac{\delta B}{B}\right)^2.
\end{equation}

Fig. 2 shows energy densities of the compressive motions and Alfv\'en
slab-type waves in two cases: first, when the CR feedback is unimportant,
but compressive motions are damped at $l_{\rm cut}$; second, when the
CR feedback become important at some scale $\lambda_{\rm fb}$. Spectral
slopes correspond to the case with $\mu=1/3$.

\begin{figure}
\includegraphics[width=84mm]{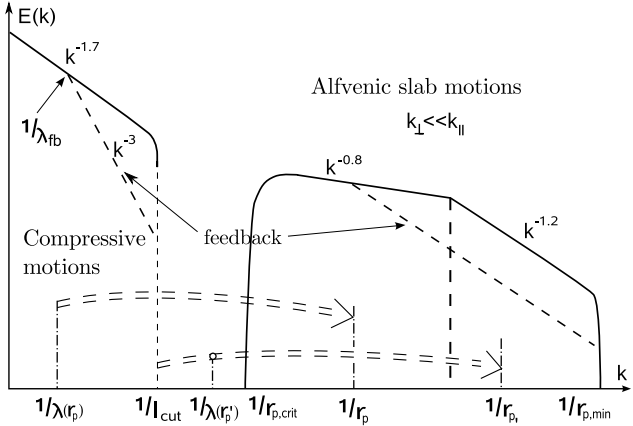}
  \caption{Energy density of compressive modes and Alfv\'enic slab-type
waves, induced by CRs. The energy is transferred from
the mean free path scale to the CR Larmor radius scale.
If the mean free path falls below compressive motions cutoff or feedback
suppression scale, the spectrum of slab waves becomes steeper.}
\label{fig1}
\end{figure}

\section{Astrophysical Consequences and Discussion}

\subsection{CRs in ISM and galaxy clusters }

Lets consider CRs in galaxy clusters. The magnetic field magnitude and the
density of CRs are somewhat uncertain there (see Esslin et al. 2005),
so we adopt values similar to our galaxy, namely $B=5\mu$G,
$n_{CR}(E>1~{\rm GeV})=4\cdot 10^{-10}~{\rm cm}^{-3}$ and $\alpha=2.6$.
This corresponds roughly to equipartition between CR and magnetic
field energies.  In clusters these energy densities are around 5 per
cent of the thermal energy density. We will have then $L_i\approx
6\cdot 10^{-7}$~pc. The reference Larmor radius of 1~GeV proton is
$r_0\approx 2\cdot 10^{-7}$~pc. We take the scale $L=1$~kpc, which,
being Alfv\'enic at this scale, corresponds roughly to driving
with the virial velocity at the scale of 30 kpc.  For these numerical
values and $\mu=1/3$ we will have, from Eq.~(\ref{amplitude}), $\delta
B/B=0.04(r_p/r_0)^{-0.1}$, almost independent on scale, $r_{p,{\rm crit}}\approx
10^{3} r_0 \approx 2\cdot 10^{-4}$~pc, $\lambda\approx 10^{-4}(r_p/r_0)^{1.2}$~pc,
and the mean free path corresponding to the turbulent
damping ($r_p=r_{p,{\rm crit}}$) is 0.4~pc which is much smaller than
the outer scale. We estimate collisionless cutoff as $l_{\rm cut}\sim 10^{-2}$~pc.
The feedback mean free path will be around 0.3 pc, so the spectrum of Alfv\'enic
slab motions will be mostly steeper, $k^{-1.2}$ and the mean free paths will be
modified according to (10) and (5). The efficient CR scattering entails efficient
second-order Fermi acceleration, see eq. (13), the process that may be important for clusters
of galaxies (Cassano, Brunetti, 2005).

In our galaxy one can assume same values for $B$, $\alpha$ and $n$ and
value of $L$ around 50~pc.  We assume an acoustic turbulence
spectrum for fast waves, taking $\mu=1/4$. We generally get a smaller
range of Alfv\'enic slab motions, from scales of $r_0$ to about
$600r_0$ with $\delta B/B=0.093(r_p/r_0)^{-0.14}$. The resulting mean free paths
$\lambda$ vary from $2.3\cdot10^{-5}$~pc to $8\cdot10^{-2}$~pc.  In
the Galactic Corona, fast waves will be damped by collisionless damping
(see, e.g., Ginzburg 1961) with a cutoff of around $1.6\cdot10^{-3}$~pc,
which is within the range of $\lambda$ that we deal with.
In the warm ionized medium (WIM) the collisional damping cutoff
will be around $10^{-4}$~pc. The feedback mean free path will be around
$7\cdot10^{-2}$. Again the spectrum of slab motions becomes steeper and
the mean free paths of CRs are modified accordingly. 

In \S~2 we assumed that the compression factor $A$ is larger than $v_A/c$.
This assumption is satisfied in galaxy clusters, as, from the
previously adopted values and $n\approx 10^{-3}$~cm$^{-3}$, $v_A/c
\approx 10^{-3}$, while compression factors for scales between 2~pc
and $2\cdot 10^{-4}$~pc are between $0.12$ and $5\cdot 10^{-3}$. For
the Milky Way ISM this condition is satisfied much better, as for
$n\approx 1$, $v_A/c\approx 4\cdot 10^{-5}$, and compression factors
are generally larger, due to the fact that the minimum $\lambda/L$
is smaller.

The slab Alfv\'en modes had been a part of the CR paradigm from the very
start of the research in the field (see Jokipii 1966). Together with
anisotropic components they are part of some of the modern 
models of CR propagation (see Zank \& Matthaeus 1992, Bieber et
al. 1994, Shalchi et al. 2006). In our model the slab
plane-parallel modes emerge
naturally as the result of the interaction of compressible turbulence
with CR. Although this mechanism is different from the earlier
considered processes, it may justify some of the earlier calculations
invoking slab modes. Unlike earlier theories we predict the dependence
on the amount of the slab mode energy on the relative pressure of the CRs.

As we see, in both clusters of galaxies and ionized gas in Milky Way the
instability within CR fluid limits the CRs mean free path. Like in scenario
discussed in YL04, where the compressible fast modes were identified as the
major CRs scattering agent, compressive modes are essential for scattering.
However in this treatment, unlike YL04, we show that compressions at
scales much larger than the resonance scale are important.
This difference is crucial for scattering of low-energy CRs, as the fast mode
have collisional or collisionless cut-offs which, depending on the media,
may be larger than the low-energy CR gyroradius. In this case YL04 appealed
to Transient Time Damping (TTD) processes, which are less efficient for
scattering than gyroresonance\footnote{In fact the gyroresonance instability
can be the major source of isotropization during the TTD acceleration.}.
Our present work shows that the slab Alfv\'en
mode discussed in the present paper can be responsible for efficient scattering.
Another important difference from YL04 is that our new mechanism require
relatively large total pressure of CRs (see (\ref{lili})). In the case
when the pressure of CRs is negligible the fast modes could stay the major
scattering agent (see Petrosian et al, 2006).

Our model predicts rather small mean free paths, but this does not contradict
the estimates on the average lifetime of the CR in the Galaxy. These lifetimes
are estimated to be around Galaxy thickness divided by the Alfv\'en velocity,
which is a powerful support for the models with streaming instability. Our model
will predict similar lifetimes, because it includes turbulence which advects
CRs on outer scale comparable with Galaxy thickness with velocity of around
Alfv\'en velocity. In fact, the turbulence itself could be generated on these
scales by Parker instability.

\subsection{Partially Ionized Gas}

Previous discussion is also applicable to
partially ionized gas, if the degree of ionization is larger than
$\sim 90\%$. Indeed, for such high ionization degrees the Alfv\'enic
turbulence cascades to scales less than the ion-neutral decoupling
scale (see Lithwick \& Goldreich 2001).

If, on the other hand, the degree of ionization is lower, we
assume that Alfv\'enic turbulence is fully damped by ion-neutral
collisions at the scale $l_{\rm damp}$, and it would not be able to provide
turbulent damping for $k_\perp l_{\rm damp}<1$. As
we saw in \S~5 the damping for slab waves with $k_\|$ is provided by
turbulent eddies with $k_\perp \sim k_\|^{3/4}$, therefore, our
slab-type component arising from CR instability will protrude up to
scales as large as $l_{\rm damp}^{4/3}/L^{1/3}$. This scale could be
substantially larger than the $r_{p,{\rm crit}}$ derived in \S~5.

According to Lazarian, Vishniac \& Cho (2004) the regime of
viscosity damped turbulence emerges for Alfv\'enic turbulence at
scales less than $l_{\rm damp}$.  This regime is characterized by a
shallow $k^{-1}$ spectrum of magnetic perturbations and it persists
down to the ion-neutral decoupling scale where it reverts to
intermittent Alfv\'enic turbulence that involves only ions. The
detailed treatment of the interactions of CRs with turbulence in
partially ionized medium is beyond the scope of this paper, however.

\subsection{Thermal plasma mean free paths in galaxy clusters}

In the paper above we considered the CR component of the ISM or ICM,
which are the high energy particles that interact with the rest of the
medium via the magnetic fields. These particles have a power-law
distribution that arises from the acceleration terms that are
proportional to the CR momenta. The astrophysical plasma, on the other hand,
is assumed to have a Maxwellian
distribution and provides us with both conductivity and mass density
which are required for a MHD treatment.  In a fully ionized plasma
particle-to-particle collisions are Coulomb scattering and the rate of
the collisions becomes smaller with temperature. With high temperature
and small density these mean free paths can be huge. For example, in
galaxy clusters it could be as large as 4kpc. This lead to apparent
contradiction, as particles with such a huge mean free path will be
subjected to acceleration and will not be Maxwellian.

Schekochihin and Cowley (2005) proposed that thermal particles
will be scattered by instabilities. They considered hydrodynamic
as well as kinetic instabilities and considered the evolution
of the cluster from initial state with no magnetic field.
Their argument is that the Reynolds number, being initially
very low, will increase with increasing magnetic field and the dynamo
will self-accelerate. They predict folded magnetic fields due
to high-Prandtl number dynamo and their mean free paths are between
viscous scale and the reversal scale.

In this subsection we estimate mean free paths of thermal particles
in a way similar to the rest of our paper, keeping in mind that
there are quite a few other plasma effects and some MHD dynamo
effects that might be important, so that these estimates are still
rather speculative. This may be excused by the fact that thermal
mean free path, viscosity and thermal conductivity are very important
for cluster dynamics.

Anisotropic distributions of thermal particles will excite waves
with inverse wavevectors of the order of thermal Larmor radius,
$r_T\approx 10^{-9}$~pc
as the instability is exponentially slow for smaller wavevectors
(see Mikhailovskii 1975, eq. 10.7). All particles will have
approximately the same mean free path, and the value of $\delta B/B$
that provides scattering will now refer to the {\it total} perturbed
magnetic field, in contrast with its definition in \S~4. Apparently
the energy-transfer arguments of \S 6 will be most important, as
the steepening is very fast on thermal Larmor scales. By equating steepening
and turbulent energy transfer rates we have $r_T/L\approx(\delta B/B)^4$, which
gives $\delta B/B\approx 10^{-3}$, $\lambda\approx 10^{-3}$~pc.

\section{Summary}

All in all, in the paper above we have demonstrated that

1. Turbulent compressions of magnetic
field result in the kinetic instability of CRs that drives Alfv\'enic
perturbations of much higher frequency with wave vectors almost parallel
to the magnetic field. These Alfv\'enic perturbations efficiently scatter
and isotropize CRs.

2. The above effect is present over the limited energy range of the CRs.
The high energy cut-off is determined by the ambient Alfv\'enic turbulence.
The non-linear back-reaction via limiting the CR mean free path and
the steepening of the generated waves control the intensity of the new
slab-type Alfv\'enic component. This intensity depends on both the
amplitude of the compressible perturbations and CR pressure.

3. The presence of linear damping of compressible motions or
the strong feedback damping effect modifies the
instability and results in a slightly steeper spectrum
of generated Alfv\'enic perturbations.

{\bf Acknowledgments} We thank Pat Diamond, Ethan Vishniac and Ellen
Zweibel for useful discussions. We thank anonymous referee for very
useful suggestions. AL acknowledges the NSF grants
AST-0307869, ATM-0312282 and the support from the Center for Magnetic
Self-Organization in Laboratory and Astrophysical Plasmas. AB thanks
IceCube project for support of his research.

\onecolumn

\appendix

\section{Cyclotron instabilities of cosmic rays in plasma}

We follow the standard procedure of deriving the dispersion relations of
electromagnetic waves in plasma. The field of the wave creates a
perturbation $f_1$ in the particle distribution function $f_0$.  We
define the current density of the perturbation as $j_i=\sigma_{ik}E_k$
where $\sigma_{ik}$ is the conductivity tensor and

\[
\epsilon_{ij}=\delta_{ij}+\frac{4\pi i}{\omega}\sigma_{ik}
\]

is a dielectric tensor. The perturbation eigenmodes are determined by
the so-called dispersion equation

\[
\left | \epsilon_{\alpha\beta}-\left(\frac{ck}{\omega}\right)^2\left(\delta_{\alpha\beta}-
\frac{k_\alpha k_\beta}{k^2}\right)\right |=0
\]
 
Symmetries of the dielectric tensor are determined by symmetries of
the initial particle distribution function. In our treatment we
consider a two-component medium in which most of the contribution into
the dielectric tensor comes from the thermal isotropic plasma, while the small
contribution from CRs is responsible for the instability. Let us consider
transverse, circularly polarized waves with wavevector parallel to the
magnetic field.  The dispersion relation will reduce to
$\epsilon_{11}\pm i\epsilon_{12}=c^2k^2/\omega^2$, and it has a large
component $e_{\pm}^{(0)}=\epsilon_{11}^{(0)}\pm i\epsilon_{12}^{(0)}$
that comes from the contribution of thermal plasma and the small component
$e_{\pm}^{(1)}$ that comes from CRs. For hydrodynamic waves with
wavelengths much larger than the Larmor radius in thermal plasma
we have $e_{\pm}^{(0)}=c^2/v_A^2$, where $v_A$ is the Alfv\'enic
velocity. Such waves propagate along the magnetic field with velocity
$v_A$. In thermal isotropic plasma both circular polarizations of the
wave have the same speed, in other words, linearly polarized waves are
also eigenmodes in such plasma. With the introduction of a CR contribution
to the dielectric tensor this degeneracy could be broken. For example, for
the isotropic distribution of CRs, ``shifted'' from the origin by some
streaming velocity, the eigenmodes are linearly polarized waves which
are unstable if the streaming velocity is larger than the Alfv\'enic
velocity (see, e.g., Kulsrud \& Pearce, 1969).  If the distribution is
not shifted, but oblate or prolate the eigenmodes are the circularly
polarized waves which are either stable or unstable (Mikhailovskii,
1975, Kulsrud 2004).

In this paper we study the oblate or prolate distributions of CRs
which come from the conservation of adiabatic invariant for
collisionless particles. The CR component $e_{\pm}^{(1)}$ is
responsible for a small imaginary part in the solution of the
dispersion relation for $\omega$. Depending on its sign the
instability constitutes the growth or damping. It could be shown that
in the limit of $v_A/c\ll1$ this growth rate is equal to
\footnote{Speaking more quantitatively, the degree of anisotropy
  should be larger than $v_A/c$ for instability to take place (Kulsrud 2004).
  In the astrophysical section we show that in the typical setting of the ISM or
  ICM our approximation is accurate enough (see \S~6).}

\[
\gamma_{\rm CR} = \pi^2e^2v_A\int \frac{v_\perp^2}{c^2}
\left(\frac{\partial F}{\partial p_\|}-
\frac{v_\|}{v_\perp}\frac{\partial F}{\partial p_\perp}\right)
\delta(k_{\|}v_\|\pm\omega_C)\,d^3{\bf p}
\]

 where
$\omega_C=eB/mc\gamma$ is a particle gyration frequency and
$F(p_\|,p_\perp)$ is a distribution function of CRs. The $\pm$ sign
correspond to the two MHD modes. We shall limit ourselves to to the
plus sign as explained in \S~3.

We introduce a small anisotropy factor A and the unperturbed
distribution function $F_0$ as

\[
A=\frac{p_\perp-p_\|}{p_\|},~~~
F_0(p)\sim\left(p_\|^2+p_\perp^2\right)^{-\alpha/2-1},
\]

where $\alpha$ is introduced in \S~2. We assume $\alpha$ to be between
2 and 3 as for the CR distribution in our galaxy. The oblate distribution
then will be described as 

\[
F(p)\sim\left(p_\|^2+p_\perp^2(1-A)^2\right)^{-\alpha/2-1},
\]

and we can, in the linear order to $A$, calculate that

\[
p_\perp\frac{\partial F}{\partial p_\|}-p_\|\frac{\partial F}{\partial p_\perp}=(-\alpha-2)AF. 
\]

Now the expression for the instability rate will be

\[
\gamma_{\rm CR} = \frac{\pi^2e^2n_{CR}}{m}\frac{v_A}{c}\frac{(-\alpha-2)A}{\omega_C}
\int \frac{\omega_C}{n_{CR}}\frac{v_\perp}{c}
F_0\delta\left(\frac{k_{\|}p_\|}{m}+\omega_C\right)\,d^3{\bf p},
\]

where we replaced $F$ with $F_0$ and introduced the cyclotron frequency
$\omega_C$, and the total density of CRs $n_{CR}$. The integral in
this expression is dimensionless.  The total density of CRs is mostly
determined by the low-energy cutoff of the distribution $F_0$ and is
rather irrelevant for the instability where only resonant particles
contribute.  It is more useful
to introduce the number of fast particles $n_{\rm
  CR}(p>m\omega_B/k_{\|})$ which is determined by the integration of $F_0$
over a region with momentum larger than the resonant momentum. After
taking the integrals we denote the gamma function as $\Gamma$ and arrive at

\[
\gamma_{\rm CR}(k_{\|}) = \omega_{pi} \frac{n_{\rm CR}(p>m\omega_B/k_{\|})}{n}AQ,
~~\mbox{where}~~~~
Q=\frac{\pi^{3/2}}{32}(\alpha+2)(\alpha-1)\frac{\Gamma(\alpha/2)}{\Gamma(\alpha/2+3/2))}.
\]



\end{document}